\documentclass[aps,prb,twocolumn,groupedaddress,amsmath,amssymb,floatfix,showpacs,citeautoscript,longbibliography]{revtex4-2}
\usepackage{graphicx}
\usepackage{dcolumn}
\usepackage{bm}
\usepackage{hyperref}
\usepackage[mathlines]{lineno}
\numberwithin{equation}{section}
\usepackage{graphicx}
\usepackage{physics}
\usepackage{amsfonts,amssymb,amsmath}
\usepackage{float}
\newcommand{\bl}{\boldsymbol}
\newcommand{\be}{\begin{equation}}
\newcommand{\ee}{\end{equation}}
\newcommand{\beqa}{\begin{eqnarray}}
\newcommand{\eeqa}{\end{eqnarray}}
\newcommand{\mr}{\mathrm{}}
\usepackage{siunitx}
\usepackage{xcolor}

\begin{document}

\preprint{}

\title{Emergence of strain-induced moir{\'e} patterns and pseudo-magnetic field confined states in graphene}

\author{Md Tareq Mahmud and Nancy Sandler}
\affiliation{Department of Physics and Astronomy, Ohio University, Athens, Ohio 45701, USA}

\date{\today}
\begin{abstract}
The mechanism by which strain-inducing deformations alter charge distributions in graphene suggests a new method to design specific features in its band structure and transport properties. Besides standard approaches used to produce distortions, novel efforts implement engineered substrates aimed to induce specifically targeted strain profiles. Motivated by this approach, we study the evolution of charge distributions with an increasing number of out-of-plane deformations aimed to mimic periodic substrates. We first analyze a system of two overlapping deformations and determine the quantitative relation between geometrical parameters and features in the local density of states. We extend the study to sets of 3 and 4 deformations in linear and two-dimensional arrays and observe the emergence of moir{\'e} patterns that are more pronounced for a hexagonal cell composed of 7 deformations. A comparison between the induced strain profile and spatial maps of the local density of states at different energies provides evidence for the existence of states confined by the pseudo-magnetic field in bounded regions, reminiscent of quantum dots structures. Due to the presence of these states, the energy level scaling to be observed by local probes should exhibit a linear dependence with the pseudo-field, in contrast to the expected scaling of pseudo-Landau levels.
\end{abstract}

\keywords{Suggested keywords}
\maketitle

\section{Introduction}

Deformations in 2D materials\cite{Dai2019} are a widespread phenomenon observed in almost all experimental setups. In graphene, substrate lattice mismatch\cite{Leroy2019,Jaroslav2019,Schouteden2015,Sule2014,GrNiReview,Stradi2012,Shu2012,Sutter2009,Wintterlin2009,Jiang2017,Mao2020,Andelkovic2020,ChenZ2014}, trapping of intercalated impurities during deposition\cite{GrCuBubbles,Briggs2019,Hwang2018,Shin2016,YWu2017,Ghorbanfekr2017,Lu2012,Jia2019}, liquid interfaces\cite{Belyaeva2020}, or purposely engineered substrates\cite{Redinger2013,Palinkas2016,NemesIncze2017,Zhang2018,Zhang2019,Pacakova2017,Reserbat2014,Stroscio2014} are the most common mechanisms responsible for their occurrence. 

Because deformations induce strain, they have a direct effect on the charge density distribution and thus provide a useful knob to modify and control electron dynamics. The comprehensive theoretical and experimental work on graphene membranes with local deformations carried on in recent years has resulted in quantitative relations between the geometry of deformations and the charge distributions produced by the underlying strain fields\cite{MVozmediano2010,SMChoi2010,NeekAmal2012, deJuan2012,Moldovan2013,Sloan2013,Ramon2014,MSchneider2015,OlivaLeyva2015,Cavalcante2016,Settnes2016,Milovanovic2016,GNaumis2017,Dawei3,Dawei2,Faria2020}. 
This correspondence provides a novel tool for sample characterization in local imaging techniques such as scanning tunneling microscopy (STM)\cite{Georgi2017,Kun2019,Li2020}.  

Isolated deformations are frequently found in supported and suspended samples -irrespective of fabrication techniques-, materializing in the form of local ripples, bubbles,  and folds. For samples deposited on clean substrates, the lattice mismatch naturally creates more extended strained regions. These  manifest as periodic structures of inhomogeneous charge distribution, known as moir{\'e} patterns, extensively studied in twisted bilayer graphene\cite{Cao2018}, a system resembling monolayer graphene on a substrate with lattice mismatch controlled by the twisting angle. 

The search for alternative protocols to create moir{\'e} patterns 'on-demand' have led naturally to the fabrication of engineered substrates with periodic structures and graphene deposited on top. Examples of these appear in recent  reports on magentotransport measurements of graphene on top of an array of insulating $SiO_{2}$ nanospheres\cite{Zhang2018,Zhang2019} and STM measurements of graphene deposited on $Au$ nanopillar arrays\cite{Jiang2017}. An important motivation for these studies is the identification of fundamental parameters that determine the pattern's characteristics and enable the design of specific band structures with novel charge transport properties\cite{Volovik2018,ChenX2014}. 

In analogy with isolated deformations, it would be useful to determine quantitative links between the geometrical parameters of  periodic structures and the characteristics of resulting moir{\'e} charge distributions. The purpose of this work is to describe the emergence of periodic structures in charge distributions of deformed graphene membranes due to repeated strain profiles induced by engineered substrates. We aim at providing quantitative relations between geometrical parameters of deformation patterns and resulting charge re-distributions as detected by local probes. 

To address these issues, we carry out a systematic study of the local density of states (LDOS) for a membrane with an increasing number of out-of-plane deformations. Our work begins with a detailed analysis of two overlapping deformations and proceeds by expanding their number one by one to analyze the effect of spatial symmetries in two distinct geometrical arrays.  

For the set of  two bubble-like out-of-plane deformations, we analyze in detail the changes per sublattice LDOS in terms of inter-bubble distance, crystalline orientation, and geometrical parameters. The study is extended to a set of 3 and 4 deformations placed in a linear array, as well as to triangular, rhomboidal, and closed pack cells formed by 3, 4, and 7 local deformations. In all these cases, we observe enhancements and depletions in the LDOS that are more pronounced in regions where the deformations overlap. The changes in the LDOS suggest charge confinement behavior reminiscent of quantum dots, with an energy level scaling that exhibits a crossover between pseudo-magnetically confined states\cite{Egger2007} and emerging pseudo-Landau levels\cite{CastroNeto2009}. Remarkably, current experimental setups make use of substrates that produce these bubble-like deformations with corrugations and widths within a $15-20 nm$ range and maximum strain of up to $2\%$\cite{Zhang2018}, resulting in typical confinement energies of the order of $50-100 meV$, suggesting that these effects should persist at room temperature. Moreover, our results reveal the emergence of periodic charge patterns {\it and establish quantitatively} their dependence with parameters of deformation profiles, thus providing a template for the design of substrates that would render desired moir{\'e} structures.

The organization of the paper is as follows: Section~\ref{sec2} presents the theoretical background used to calculate changes in LDOS, based on the continuum (Dirac) model for electrons in graphene and continuum elasticity theory. In Section~\ref{sec3} this formalism is applied to a two-bubble system with identical and different geometrical parameters. Section~\ref{sec4} presents numerical results for linear arrays of 3 and 4 deformations, as well as for triangular, rhomboidal, and one hexagonal array of 7 deformations.  We contrast results for these various arrangements and summarize our findings in the Conclusions section.

\section{Theoretical Background }
\label{sec2}

We consider an undistorted (pristine) graphene layer (lattice constant $a=2.46\AA$), lying on the $(x-y)$ plane, with the x-axis along the zigzag direction, and use the effective low-energy continuum model to describe the electron dynamics. The corresponding Hamiltonian is written in the valley isotropic basis\cite{Beenakker2008}:

\be
H_{0}=v_{F}\bf{\sigma} \cdot \bf{p} 
\label{eq;H0}
\ee

For the sake of completeness, we first review the standard procedure used to describe strain induced by an isolated out-of-plane deformation, modeled by a Gaussian-shaped geometry centered at position ${\bf{r}_{0}}=(x_{0},y_{0})$:
\be
h({\bf{r},\bf{r}_{0}})= h_{0} \exp(-\frac{|{\bf{r}-\bf{r}_{0}}|^2}{b^2})
\label{eq:h(r)}
\ee
with $h_{0}$ as the maximum amplitude and $b$ related to the full width at half-maximum of a Gaussian function by FWHM=$4b\sqrt{\ln{2}}$. We assume that the deformation is smooth on the interatomic length-scale and describe atomic displacements with continuum elasticity theory. For a deformation of this kind, inter-valley scattering can be neglected when $b \gg a$, hence each valley can be treated separately. The induced strain is given in terms of the strain tensor $\epsilon$ with components defined by:

\be
\epsilon_{lm}=\frac{1}{2} \bigg(\partial_{l}u_{m} + \partial_{m}u_{l}+\partial_{l}h_{i}\partial_{m}h_{i}\bigg),
\label{eq:epsilon}
\ee
where we have used the implicit sum notation in the non-linear term.

Here $u$ and $h$ stand for in-plane and out-of-plane displacements respectively. As discussed elsewhere\cite{DaweiMPLB}, non-linear terms due to out-of -plane displacements are largely responsible for the emergence of the inhomogeneous charge profiles and linear terms can be safely neglected.  When strain is incorporated in a tight-binding model, changes in nearest neighbors hopping parameters are captured by particular combinations of the strain tensor components, with opposite signs between valleys. In the continuum limit, these combinations are arranged to form the components of a pseudovector field $\bf{A}(\bf{r})$, that we chose as follows:  
\be
A_{x}=-\frac{\hbar\beta}{2ea}(\epsilon_{xx}-\epsilon_{yy}), \quad A_{y} = \frac{\hbar\beta}{ea}\epsilon_{xy}.
\label{eq:A}
\ee
Here $\beta\approx 3$ is related to the Gruneisen parameter as described in Ref.~\onlinecite{Ando2002}, and $e$ stands for the electron charge. Note that these expressions are written for valley $K$ and have opposite signs at valley $K'$. Inhomogeneities in $\bf{A}(r)$ can give rise to a pseudo-magnetic field $\bf{B}(\bf{r}) = \bigtriangledown \times \bf{A}(\bf{r})$, a quantity that is used to provide an intuitive description for changes in electron dynamics in the presence of the deformation.
Simultaneously, the trace of the tensor gives rise to a scalar field
\be
U({\bf{r}})= g_{s}(\epsilon_{xx}+\epsilon_{yy}),
\label{eq:U}
\ee
that imposes the conservation of charge neutrality within each unit cell\cite{Ando2002}. Here $g_{s}$ stands for the coupling constant with values reported in the literature within the range $2-3 eV$\cite{Midtvedt2016}. 

The electron dynamics in strained graphene is thus described by an effective Hamiltonian given by:
\be
H_{\tau}=v_{F}{\bf{\sigma}}\cdot[{\bf p} + \tau e {\bf{A}(\bf{r})}] + \sigma_{0}U({\bf{r}}).
\label{eq:total}
\ee
where $\tau = \pm1$ stands as a label for $K$ and $K'$ respectively.
In the last term, $\sigma_{0}$ is the 2$\times$2 identity matrix. 
The LDOS $\rho_{j}({\bf{r}}, E)$ defined for sub-lattice $j$, (where $j = 1, 2$ stands for $A , B$ respectively) at position $\bf{r}$ and energy E, is obtained via standard Green's functions methods:
\be
\rho_{j}({\bf{r}}, E) = -\xi \frac{1}{\pi}{\cal{I}}m\;G({\bf{r},\bf{r}},E)_{jj}.
\label{eq:LDOS}
\ee
$\xi =1 (-1)$ for $E >0 (< 0)$ and $G({\bf{r},\bf{r}},E)_{jj}$ is the diagonal element of the $2\times 2$ single particle Green's function
\be
G({\bf{r},\bf{r}'}, E)=\sum_{\tau} \bra{{\bf{r}}}\frac{1}{E-H_{\tau}+i \xi(E)0^{+}}\ket{{\bf{r}'}},
\label{eq:GF}
\ee
that, in perturbation theory, can be written as
\begin{equation}
G({\bf{r},\bf{r}'},E) = G_{0}({\bf{r},\bf{r}'},E)+\Delta G({\bf{r},\bf{r}'},E)
\label{eq:GFpt}
\end{equation}
where $G_{0}$ denotes the Green's function for pristine graphene and $\Delta G(\bf{r},\bf{r}',E)$ is the correction introduced by 
the Hamiltonian:
\be
H^{int}_{\tau}({\bf{r}})=\tau v_{F} e {\bf{\sigma}}\cdot{\bf{A}(\bf{r}})+\sigma_{0} U({\bf{r}}).
\label{eq:Hint}
\ee
The explicit expression for valley $\tau$ to first order in $H^{int}_{\tau}({\bf{r}})$ is simply:
\be
\Delta G_{\tau}({\bf{r},\bf{r}'},E) = \int_{{\bf{r}_{1}}} G_{0, \tau}({\bf{r},\bf{r}_{1}},E)H^{int}_{\tau}({\bf{r}_{1}}) G_{0, \tau}({\bf{r}_{1},\bf{r}'},E). 
\label{eq:deltaG}
\ee

We recall the expression for pristine graphene's Green's function at valley $\tau$, given by\cite{Katsnelson2012,Bena2009,DaweiMPLB}
\be
G_{0, \tau}({\bf{r},\bf{r}'},E)=-\frac{k}{4\hbar v_{F}}
\begin{pmatrix}
i \xi H^{(0)}(kd) & -\mathrm{e}^{-i\theta}H^{(1)}(kd)\\
-\mathrm{e}^{i\theta}H^{(1)}(kd) & i \xi H^{(0)}(kd)
\end{pmatrix}
\label{eq:G0}
\ee
where $H^{(0)}(kd)$ and $H^{(1)}(kd)$ are the first and second order Hankel functions of the first kind respectively, $k=|E|/\hbar v_{F}$, and $\bf{d}=\bf{r}-\bf{r}_{1}$. From Eq.~\ref{eq:G0}, the LDOS for pristine graphene per sub-lattice, spin orientation and unit cell area is:
\be
\rho_{0}=\frac{|E|}{2\pi\hbar^{2}v^{2}_{F}}
\label{eq:ldos0}
\ee  

Using Eqs.~\ref{eq:Hint}, \ref{eq:deltaG}, and \ref{eq:G0} we obtain the sub-lattice LDOS for deformed graphene:
\be
\rho_{j}({\bl{r}},E) = \rho_{0}(E)+ \Delta \rho_j({\bf{r}},E)
\label{eq:ldosj}
\ee
where $\Delta \rho_j({\bf{r}},E)= \displaystyle\sum_{\tau}\Delta\rho_{\tau, j}({\bf{r}},E)$ is the change in the $j$-sub-lattice LDOS produced by the deformation. 

For the specific Gaussian deformation described by Eq.~\ref{eq:h(r)}, the expressions for the pseudo-vector and scalar fields in Eqs.~\ref{eq:A} and~\ref{eq:U} are:

\beqa
U({\bf{r}}) &=&\eta^2 g_{s}\frac{R^2}{b^2} e^{-\big(2\frac{R^2}{b^{2}}\big)} \label{eq:Ububble}\\
\bf{A}(\bf{r}) &=&\eta^2 \frac{g_{v}}{ev_{F}}\frac{R^{2}}{b^2}e^{-\big(2\frac{R^2}{b^2}\big)}(-\cos2\gamma,  \sin2\gamma)
\label{eq:Abubble}
\eeqa
Here $v_{F}$ is the Fermi velocity, $g_{v}=\hbar\beta v_{F}/2a \approx 7eV$ is the coupling strength for the pseudo-vector field, $R=|\bf{r}-\bf{r}_{0}|$, and $\gamma$ as the angle between $\bf{R}$ and the $x$-axis. We introduced the parameter $\eta = (h_0/b)$ as a measure of the strain strength in the Gaussian deformation model. The use of continuum elasticity theory for this kind of deformation is valid when $\eta \ll 1$\cite{MVozmediano2010}.  

\section{Two local deformations}
\label{sec3}

We proceed to discuss the effects of two identical Gaussian bubbles of height $h_{0}$ and width b, centered at positions $\bl{r_{1}}=(x_{1}, y_{1})$ and $\bl{r_{2}}=(x_{2}, y_{2})$ along the zigzag crystalline orientation, and separated by $\bl{d} = \bl{r_1} - \bl{r_2}$. In this arrangement, the shape of the membrane is described by the function
\be
h({\bf{r}})= h_{0} \sum_{i = 1,2}\exp(-\frac{|{\bf{r}-\bf{r}_{i}}|^2}{b^2}), 
\label{eq:htwobubbles}
\ee
Clearly, the global rotational symmetry present for an isolated deformation is  broken, and the expressions for the strain induced fields are correspondingly modified. The scalar potential is written as
\begin{equation}
U(\bl{r}, \bl{r_{1}}, \bl{r_{2}})= \sum_{i=1,2}U({\bl{r}, \bl{r_i}}) + U_{ov}(\bl{r}, \bl{r_{1}}, \bl{r_{2}}) 
\label{eq:totalU}
\end{equation}
where $U({\bl{r},\bl{r_i}})$ correspond to the potentials produced by each individual deformation (Eq.~\ref{eq:Ububble}), and $U_{ov}(\bl{r}, \bl{r_{1}}, \bl{r_{2}})$ stands for the potential in the overlap region:
\begin{equation}
U_{ov}(\bl{r},\bl{r'}) = 4g_{s}\eta^2 \bigg(\frac{|\bl{r}-\bl{r'}|^{2}}{b^2}-\frac{d^2}{2b^2}\bigg)\mr{e}^{-(2|\bl{r}-\bl{r'}|^{2}+d^2)/b^2}
\label{eq:Uov}
\end{equation}
where we have defined $\bl{r'}=\Big[\frac{(x_{1}+x_{2})}{2} \bl{i}+\frac{(y_{1}+y_{2})}{2} \bl{j}\Big] $ and $d = |\bl{d}|$. As expected, Eq.~\ref{eq:Uov} contains an exponential decay proportional to the separation between the centers of the bubbles while maintaining the rotational symmetry with respect to $\bl{r'}$.

\begin{figure}[ht]
\includegraphics[width=0.50\textwidth]{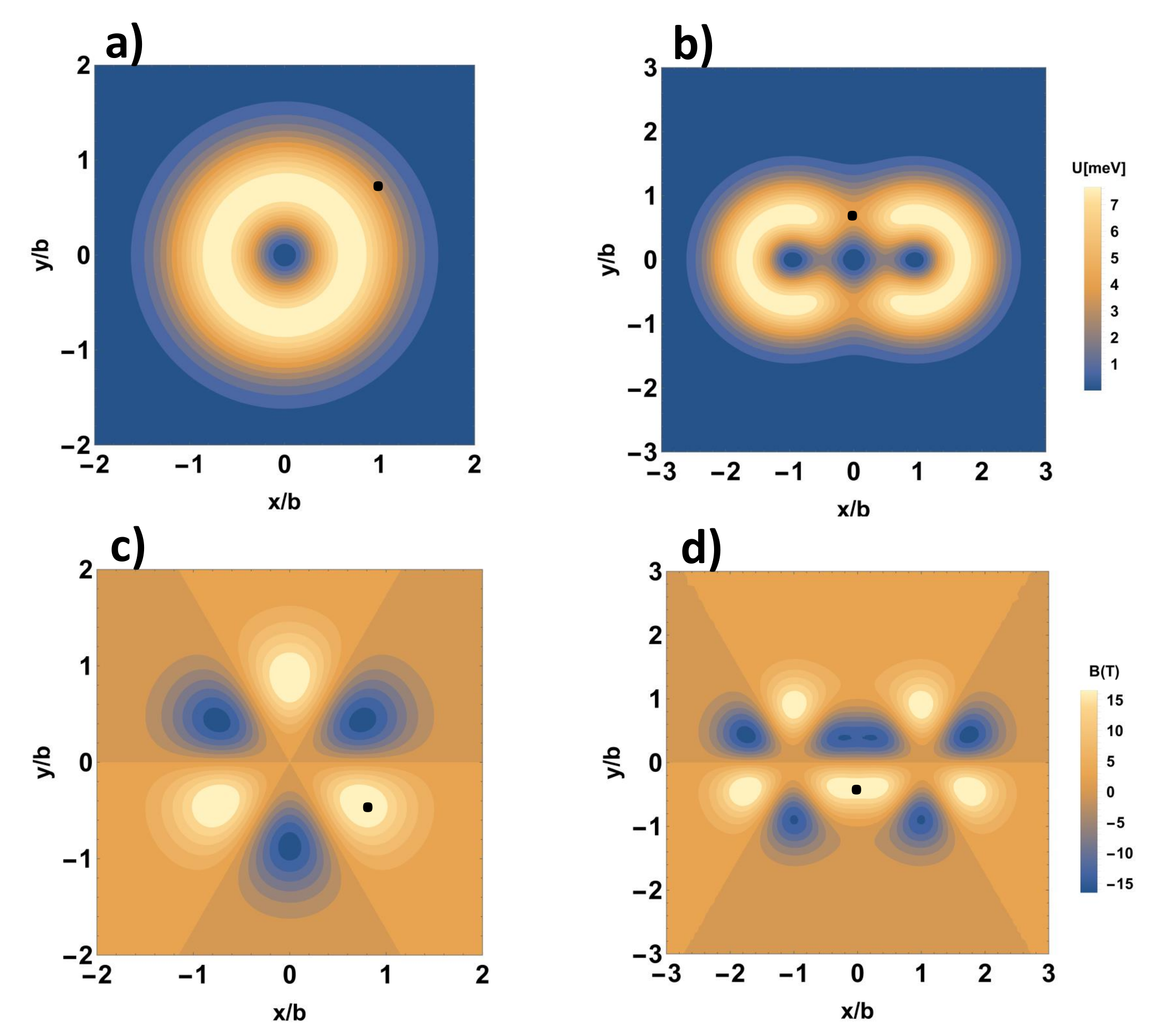}
\caption{Strain-induced field profiles for one and two deformations, at valley $K$. Scalar potential $U$: a) single bubble; b) double bubble. Results at valley $K'$ are identical and not shown. Pseudo-magnetic field $B$: c) single bubble; d) double bubble. Results for valley $K'$ show reversed color at each region. Parameter values: $h_{0}=1nm$, 
$b=10nm$, $g_{s}=2eV$ and $g_{v}=7eV$. For the double bubble system, centers are located at $(-b,0)$ and $(b,0)$ respectively. Black dots signal positions for plots shown in Fig.~\ref{fig:Fig2}.}
\label{fig:Fig1}
\end{figure}

Similarly, the total pseudovector potential takes the form:
\be
\bl{A}(\bl{r}, \bl{r_{1}}, \bl{r_{2}})=\sum_{i=1,2}\bl{A}(\bl{r},  \bl{r_{i}}) + \bl{A_{ov}}(\bl{r}, \bl{r_{1}}, \bl{r_{2}}).
\label{eq:totalA}
\ee
The expression for the overlap term at valley $K$ is given by:
\be
\bl{A_{ov}}=\frac{g_v\eta^2}{ev_F}\bigg(\frac{d_{1}^{2}}{b^2}\bigg)\bigg(\frac{d_{2}^{2}}{b^2}\bigg)\mr{e}^{-(d_{1}^{2}+d_{2}^{2})/b^2}
\begin{pmatrix}
-\frac{1}{2}\cos(\gamma_{1}-\gamma_{2})\\
\sin(\gamma_{1}+\gamma_{2})
\end{pmatrix},
\label{eq:Aov}
\ee
where $\bl{d_{1}}=\bl{r}-\bl{r_{1}}$ and $\bl{d_{2}}=\bl{r}-\bl{r_{2}}$. $\gamma_{1}$ and $\gamma_{2}$ are the angles formed between $\bl{d_{1}}$ and $\bl{d_{2}}$ and the x-axis respectively. Figure~\ref{fig:Fig1} shows plots for the scalar potential and pseudo-magnetic fields derived from $\bl{A}(\bl{r})$ for one and two overlapping bubbles with centers at $(-b, 0)$ and $(b, 0)$, a separation that allows to identify each bubble individually. Characteristic features of the fields produced by each individual deformation can be identified away from the center, while an extended structure emerges in the region between them, as a result of their overlap. While the profile for the total scalar field exhibits the rotational symmetry around the center described by Eq.~\ref{eq:Uov}, the final shape of the pseudo-magnetic field depends on the orientation of the line that joins the pair of deformations with respect to the crystalline axis. Panel d) illustrates the field for an orientation parallel to the zigzag direction, produced by the combination of fields of isolated deformations with the same sign in the region of overlap. The combination of individual fields in the overlap region, provides an intuitive way to visualize and predict areas with increased (or depleted) LDOS as the number of deformations is increased. Interestingly, the magnitude of the field shows small variations across this area, suggesting that overlap areas may sustain quasi-constant pseudo-magnetic fields. (Signs of the pseudo-magnetic field regions are reversed at valley $K'$, while the overall pattern remains the same.)

\begin{figure}
\includegraphics[width=0.485\textwidth]{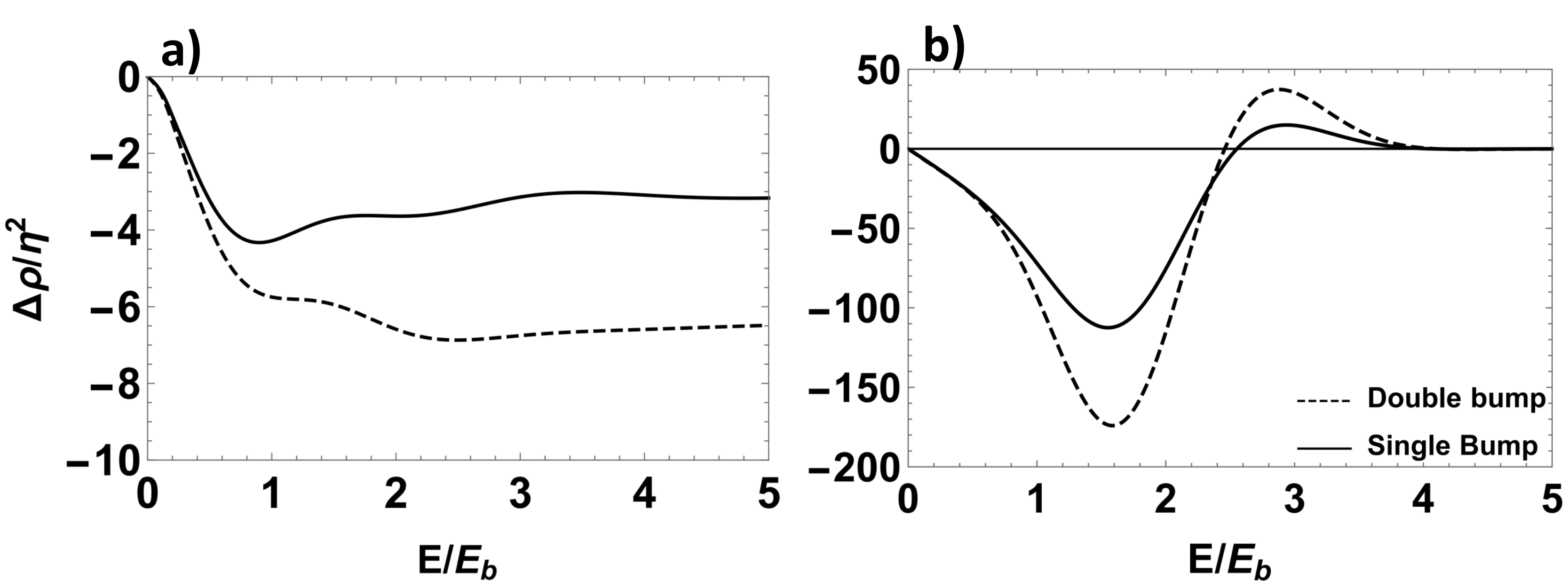}
\caption{$\Delta \rho_j({\bf{r}},E)$ profile (in units of  $10^{-4} eV^{-1}\AA^{-2}/\eta^2$), as a function of dimensionless energy $E/E_b$, for single (solid line) and double bubble (dashed line) deformations at positions marked by black dots in Fig.~\ref{fig:Fig1}. a) Variations due to scalar and b) pseudo-magnetic fields. The total variation in LDOS (not shown) follows the profile produced by the pseudo-magnetic field in panel b), as its effect is two orders of magnitude larger at the energies of interest. Parameter values: $h_{0}=1nm$, $b=10nm$, $\eta^2 = 10^{-2}$, $g_{s}=2eV$ and $g_{v}=7eV$.}
\label{fig:Fig2}
\end{figure}

Next, we implement the procedure described in the previous section to obtain the LDOS for two bubbles, as a function of energy $E$, bubble separation $d$ and geometrical parameters $h_0, b$. Results presented in the rest of the paper are given in terms of the energy scale set by one deformation, defined as:
\be
E_b = \frac{\hbar v_F}{b}.
\label{eq:Eb}
\ee  
Fig.~\ref{fig:Fig2} presents $\Delta \rho_j({\bf{r}},E)$ normalized by strain strength $\eta$ due to the scalar potential, pseudo-magnetic field and both fields combined, for a single bubble centered at $(0, 0)$ (solid line), and two bubbles centered at $(-b, 0)$ and $(b, 0)$ (dashed line) respectively. To highlight the changes in the overlap region, plots are shown for the specific positions marked in Fig.~\ref{fig:Fig1} that correspond to regions where the two deformations overlap. Note that the points indicate the same spatial location with respect to the center of a given deformation, although their coordinates are different due to the reference frames chosen in each case. Panel a) shows that changes caused by the scalar potential converge to a constant value for larger energies. This local breaking of particle-hole symmetry has the effect of a shift in the chemical potential\cite{Dawei2} at the deformation. When the second deformation is added, the value for $\Delta \rho_j({\bf{r}},E)$ shows variations of over $50\%$ (see values at $E \simeq E_b$ in panel b)), a clear signature of enhanced electron depletion in this area. Similar plots with opposite signs, i.e., showing increased LDOS, are obtained for positions located at the center of blue areas in Fig.~\ref{fig:Fig1}. The maximum change introduced by the pseudo-magnetic field is two orders of magnitude larger than the corresponding maximum change due to the scalar field and, as a consequence, the combined effects are dominated by the pseudo-magnetic field. 

\begin{figure}
\includegraphics[width=0.48\textwidth]{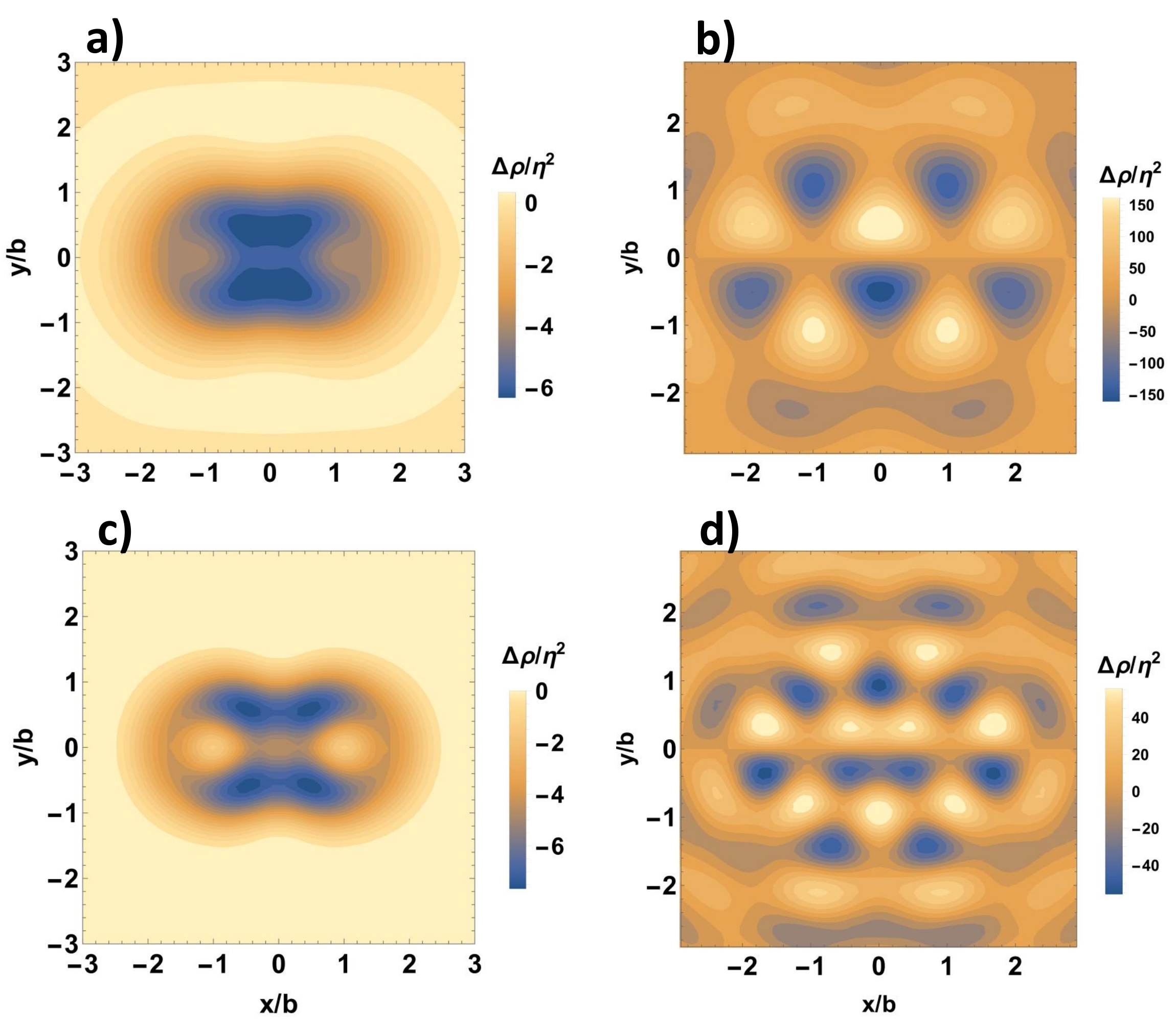}
\caption{Real space profile for $\Delta \rho_j({\bf{r}},E)$  at energies $E=1.6E_{b}$ (top row), and $E=2.5E_{b}$ (bottom row) for one and two deformations. Left and right columns show changes induced by scalar and pseudo-magnetic fields respectively. Units and other parameters as in Fig.~\ref{fig:Fig2}.}
\label{fig:Fig3}
\end{figure}

The variation of $\Delta \rho_j({\bf{r}},E)$  shown in Fig.~\ref{fig:Fig2} hints at different types of enhancement/depletion as the energy is varied. To illustrate this effect, we plot in Fig.~\ref{fig:Fig3} data for the real space distribution of the LDOS at two different energies $E=1.6E_{b}$ and $E=2.5E_{b}$. Left and right columns show results for changes induced by scalar and pseudo-magnetic field respectively. Notably, the changes induced by the scalar field exhibit similar spatial inhomogeneities at different energies while those produced by the pseudo-field are clearly different. However, as already pointed out,  their magnitudes are consistently smaller than those due to the pseudo-magnetic field. In the rest of the paper we focus only on the changes produced by the pseudo-magnetic field.

The finite size areas with increased LDOS surrounded by areas of charge depletion, shown in Fig.~\ref{fig:Fig3}b), are reminiscent of quantum dot structures. In contrast, panel d) shows extended areas with a quasi-one dimensional geometry. These results suggest that changes in the chemical potential can induce transitions between different types of carrier confinement (or depletion), with remarkably different physical behavior in transport properties.

\begin{figure}
\includegraphics[width=0.49\textwidth]{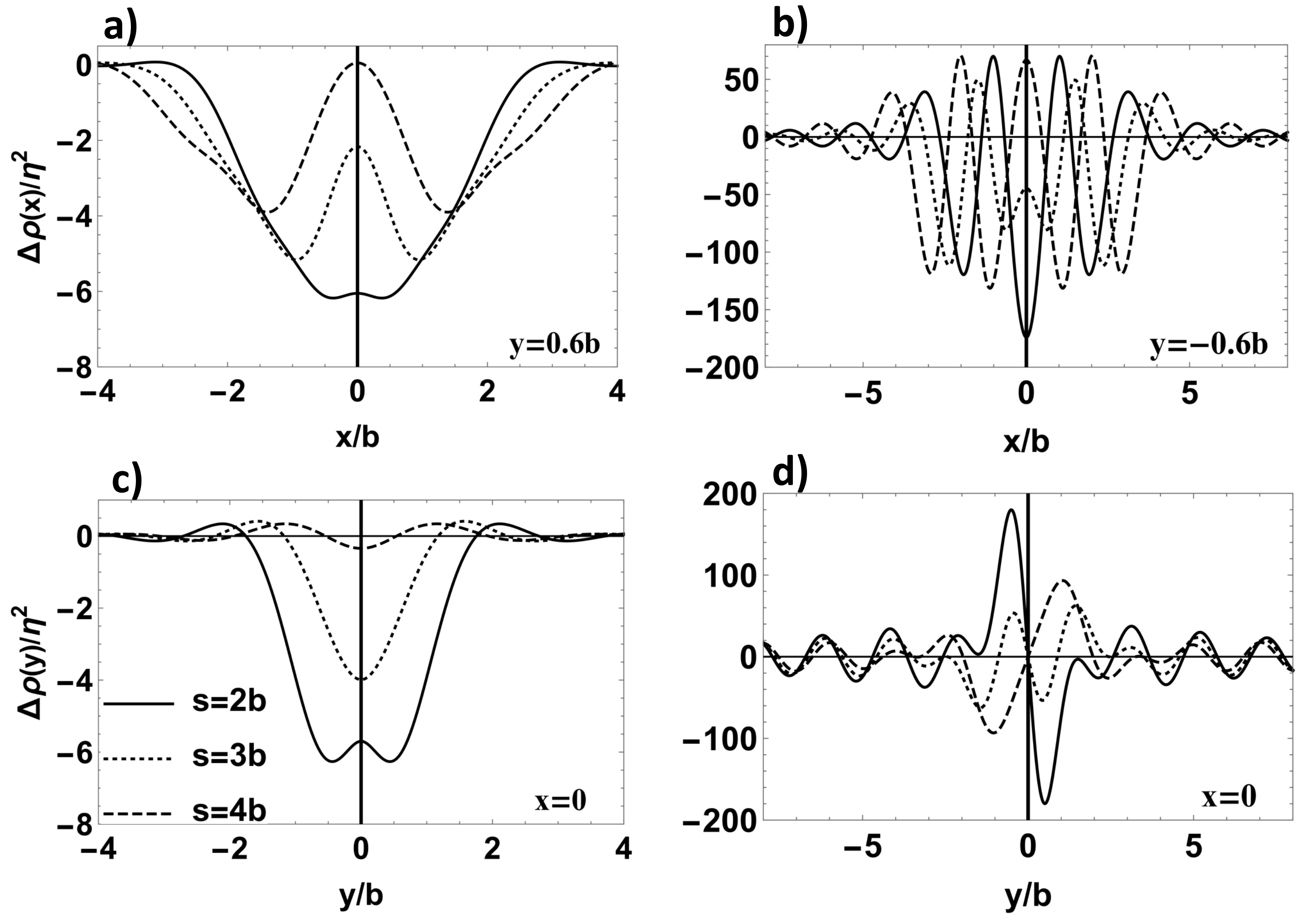}
\caption{
$\Delta \rho_j({\bf{r}},E)$  profile along the zigzag $(x)$ and C-C bond $(y)$ directions due to scalar (panels a) and c)); and pseudo-magnetic fields (panels b) and d)); for two bubbles at separations $s=2b$ (solid line), $s=3b$ (dotted line), and $s= 4b$ (dashed line), at $E=1.6 E_b$. Ranges for horizontal axis chosen to emphasize the different extension of changes produced by each strain-induced field. Units and other parameters as in Fig.~\ref{fig:Fig2}.}
\label{fig:Fig4}
\end{figure}

The dependence of the magnitude of $\Delta \rho_j({\bf{r}},E)$ with the separation between bubbles, is shown in Fig.~\ref{fig:Fig4}. Changes due to each separate field are shown for two deformations in three different configurations, at distances $s=2b, 3b$ and $s=4b$, at fixed energy $E = 1.6 E_b$. Panels a) and b) show $\Delta \rho_j({\bf{r}},E)$  along the zigzag direction (line joining the two deformations), while panels c) and d) show similar results along the C-C bond. 
The cuts shown are taken at the values of $(x,y)$ coordinates that correspond to positions of maximum overlap (and LDOS change) in each case respectively.
These plots suggest that the effects on the overlap region vanish for separations  $s \geq 4b$, and the values of LDOS return to those produced by individual bubbles.

\begin{figure}
\includegraphics[width=0.49\textwidth]{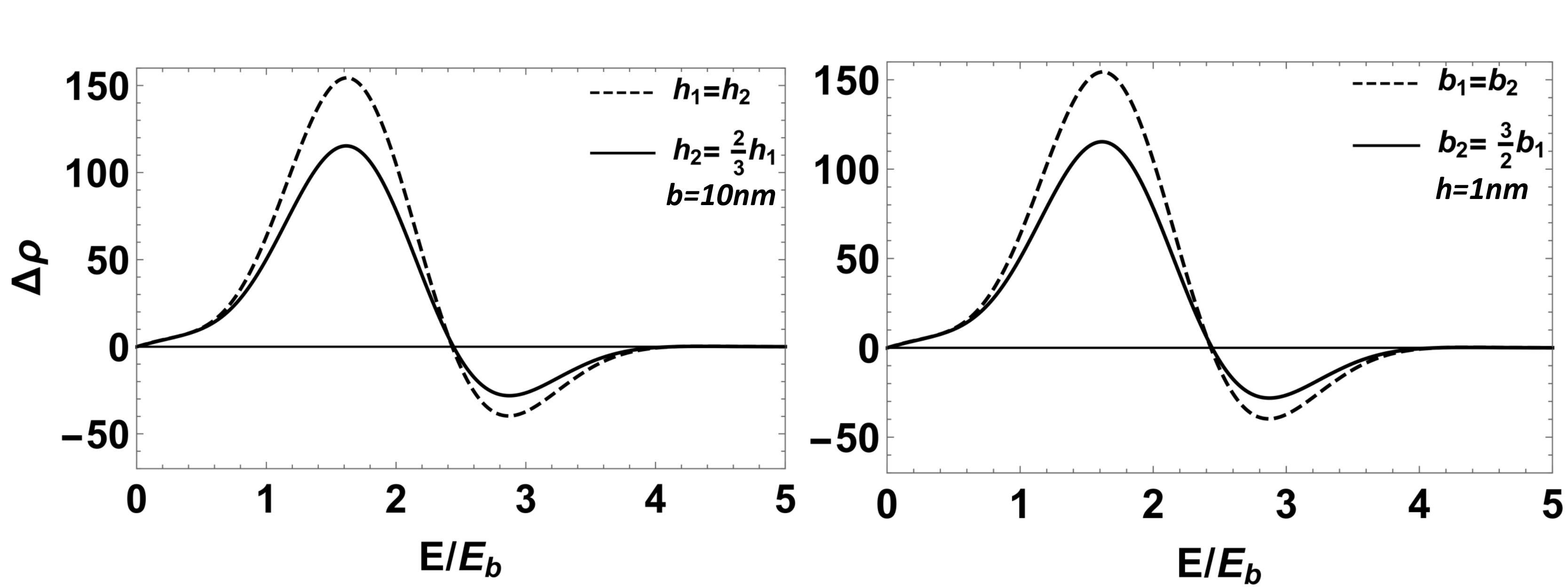}
\caption{Comparison of $\Delta \rho_j({\bf{r}},E)/\eta^{2}$ for two Gaussian deformations with identical (dashed lines) and different (solid lines) geometrical parameters at the position marked in Fig.~\ref{fig:Fig1}d). Panel a): identical half-width $b = 10 nm$ and heights $h_1 = 1.0 nm$ and $h_2= 0.67 nm$. Panel b): identical height $h = 1 nm$ and half-widths $b_1 = 10 nm$ and $b_2 = 15 nm$. In both cases $C=0.6$. Units and other parameters as in Fig.~\ref{fig:Fig2}.}
\label{fig:Fig5}
\end{figure}

The analysis carried out above considers the case of two identical deformations, a situation not generally observed in experimental settings. In order to estimate the consequences of different geometrical parameters, we analyzed data for configurations with different height $h_0$ and same value of $b$; and also with different values of $b$ and same height $h_0$. Either change involves different strain strengths, and the final shape of the membrane is determined by the deformation inducing higher strain. Let's introduce $\eta_1$ and $\eta_2$ as the new strain strengths, with $\eta_1 > \eta_2$. From Eqs.~\ref{eq:deltaG}, ~\ref{eq:totalU}, ~\ref{eq:Uov}, ~\ref{eq:totalA}, and ~\ref{eq:Aov},  
the corresponding total change can be expressed as:
\be
\Delta \rho_j = \eta_{1}^{2}\Delta \rho_{j,1} + \eta_{2}^{2}\Delta \rho_{j,2} + \eta_{1}\eta_{2}\Delta \rho_{j,ov}.
\label{eq:diffbubbles}
\ee
Here, $\Delta \rho_j({\bf{r}},E)$  stands for the change due to the individual deformation with strain $\eta_i$ and $\Delta \rho_{j,ov}$ for the change in the overlap region. It is convenient to introduce the factor $C= \eta_2/\eta_1$, with $C \leq 1$, 
\be
\frac{\Delta \rho_j}{\eta_{1}^{2}} = \Delta \rho_{j,1} + C\,\Delta \rho_{j,ov} + C^2\, \Delta \rho_{j,2}.
\label{eq:diffbubblesC}
\ee
Not surprisingly, the change in the region with smaller strain decreases quadratically with $C$ but the change in the overlap region decreases only linearly with $C$. Typical variations of geometrical parameters in experimental settings\cite{Zhang2018} give values of $C$ in the range $0.8-0.9$, hence, the variations in the overlap regions are thus expected to be within $10-20\%$. Representative results are shown in Fig.~\ref{fig:Fig5}, that plots $\Delta \rho_j({\bf{r}},E)$ for two deformations at a fixed position with different geometrical parameters. 

On a more general perspective, a substrate with similarly shaped deformations but with widely different characteristic parameters (i.e., with $C \ll 1$), should render changes in LDOS consistent with those introduced by the deformation inducing the largest strain fields. The resulting LDOS pattern then would correspond to a set of isolated deformations with minimal overlaps.  

\section{Towards a periodic structure}
\label{sec4}

Having fully characterized the properties of a two-bubble system, we now focus on an increased number of deformations to describe the crossover to a periodic structure. First, we consider one-dimensional arrays of 3 and 4 overlapping Gaussian bubbles with centers along the zigzag direction (i.e. along the x-axis). Then, we compare with triangular and closed packed-structures and discuss the emergence of moir{\'e} patterns.

\begin{figure}
\includegraphics[width=0.49\textwidth]{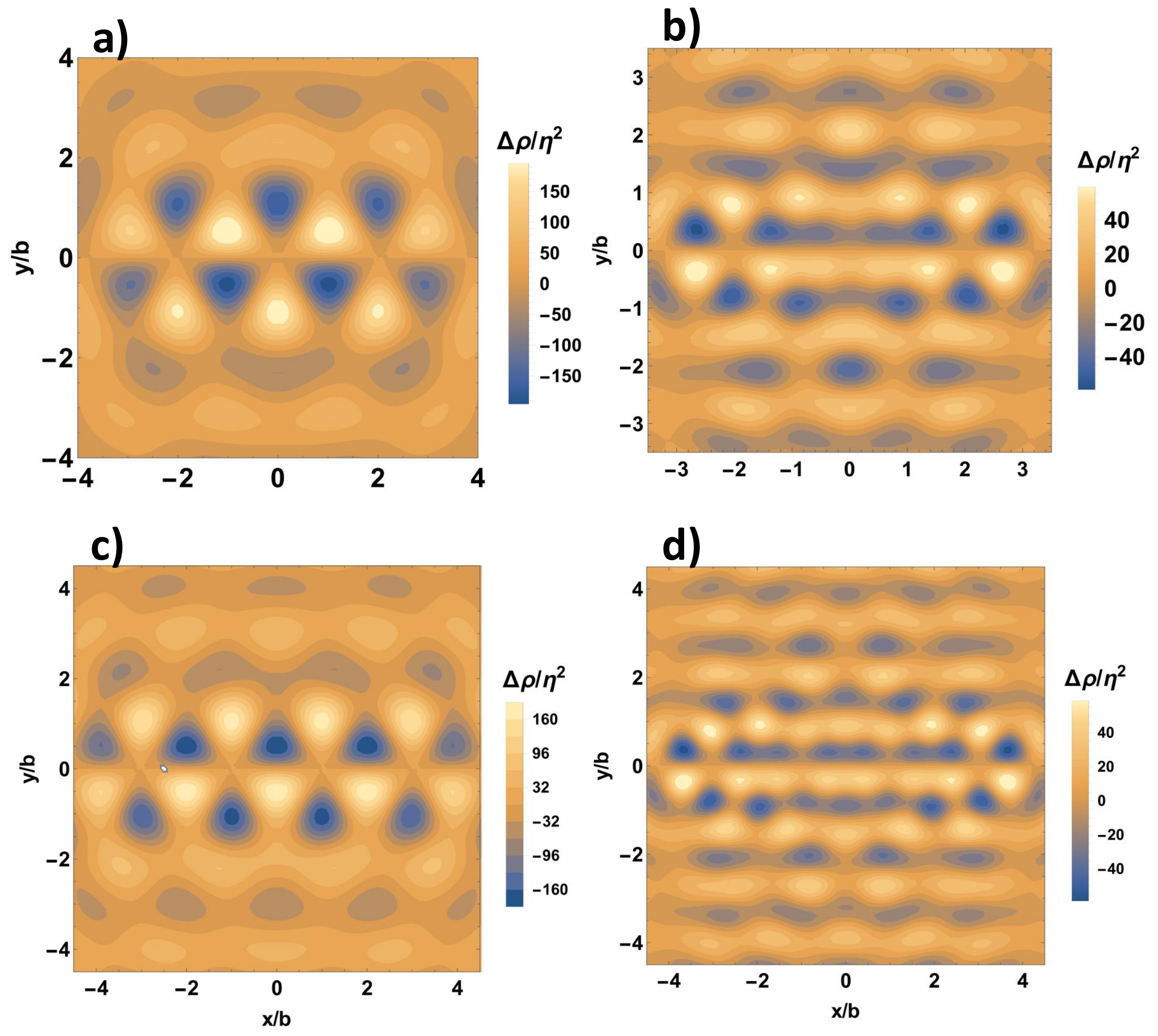}
\caption{Real space profile for $\Delta \rho_j({\bf{r}},E)$ induced by a linear array of identical bubbles. Panels a) and b): three deformations with centers at $(-2b, 0)$, $(0, 0)$, $(2b, 0)$. Panels c) and d): four deformations with centers at $(-3b, 0)$, $(-b, 0)$, $(b, 0)$, $(3b, 0)$. Left panels show results for $E=1.6 E_{b}$ and right panels for $E=2.5 E_{b}$ respectively. Units and other parameters as in Fig.~\ref{fig:Fig2}.}
\label{fig:Fig6}
\end{figure}

\begin{figure}
\includegraphics[width=0.45\textwidth]{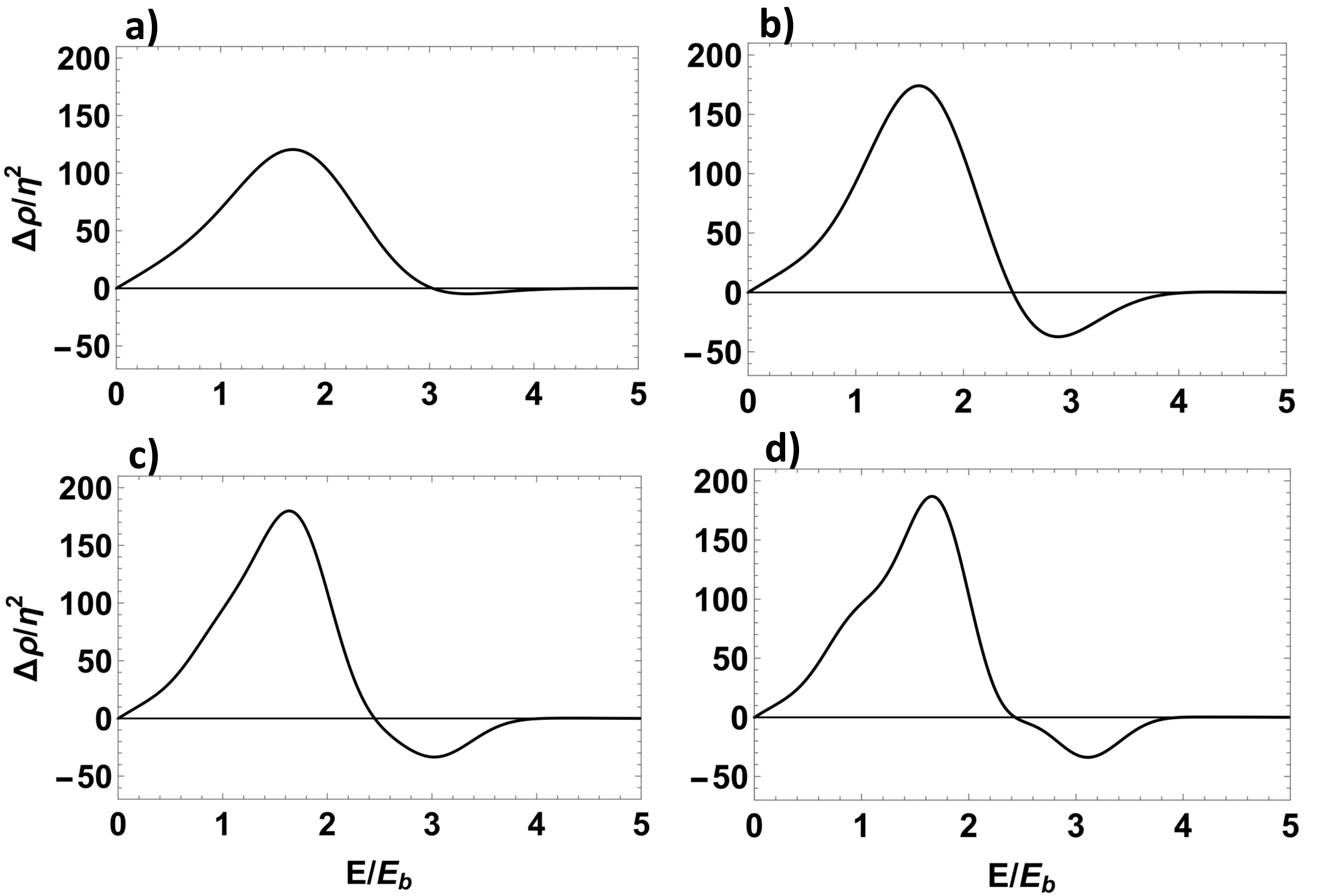}
\caption{$\Delta \rho_j({\bf{r}},E)$ as a function of $E/E_b$ for increasing number of identical deformations in a linear configuration parallel to the zigzag direction. $\Delta \rho_j({\bf{r}},E)$ is measured at real space positions corresponding to maximum magnitudes. Panel a): one bubble. Panel b) two bubbles. Panel c) three bubbles. Panel d) four bubbles. Units and other parameters as in Fig.~\ref{fig:Fig2}.}
\label{fig:Fig7}
\end{figure}

Fig.~\ref{fig:Fig6} presents real space profiles for $\Delta \rho_j({\bf{r}},E)/\eta^2$  at energies $E=1.6 E_{b}$ (left panels), and 
$E=2.5 E_{b}$ (right panels) for three (top row) and four (bottom row) bubbles in a linear array. In general, regions with maximum changes in LDOS become better defined as the number of deformations is increased. Right panels show larger confinement (better seen in Fig.~\ref{fig:Fig7} discussed below) in well separated areas. In contrast, left panels show the emergence of parallel quasi-one dimensional regions with structure resembling sets of parallel wires.

To measure the evolution of $\Delta \rho_j({\bf{r}},E)$ with the number of deformations we plot in Fig.~\ref{fig:Fig7} cross cuts taken at positions of maximum intensity for one bubble and arrays of two, three and four bubbles. 
These plots show an increase in the overall magnitude of $\Delta \rho_j({\bf{r}},E)$ with the number of bubbles, and a better defined peak in a narrow range of energies. Although the maximum change occurs when adding a second deformation (due to the direct overlap between regions with same pseudo-magnetic field signs, as discussed in the previous section), the addition of more bubbles produce further increases in the magnitude of $\Delta \rho_j({\bf{r}},E)$ and defines a much sharper peak. These increases can be seen as due to smaller but not negligible contributions described by the long-range oscillations in the Green's function (see Eqs.~\ref{eq:deltaG} and ~\ref{eq:G0}). 

\begin{figure}
\includegraphics[width=0.49\textwidth]{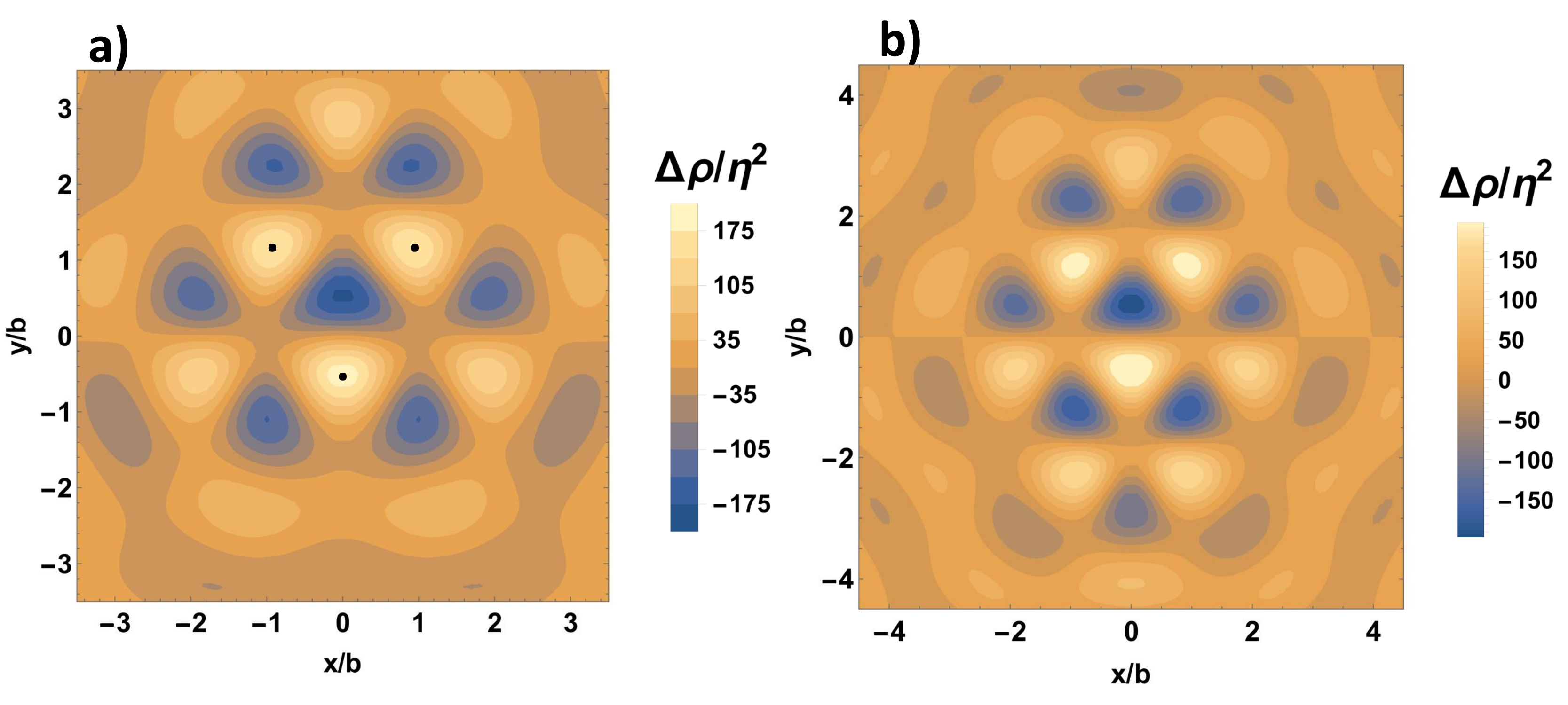}
\caption{Real space profile for $\Delta \rho_j({\bf{r}},E)$ for three and four identical bubbles in triangular and rhomboidal configurations.  Panel a) bubble centers located at $(-b,0)$, $(b, 0)$, $(0, \sqrt{3}b)$. Panel b) bubble centers located at $(-b,0)$, $(b, 0)$, $(0, \sqrt{3}b)$ and $(0, -\sqrt{3}b)$. Data plotted for $E=1.6 E_{b}$. Black dots indicate positions for plots in Fig.~\ref{fig:Fig9}. Units and other parameters as in Fig.~\ref{fig:Fig2}.}
\label{fig:Fig8}
\end{figure}

A similar analysis can be made for two-dimensional configurations formed by three and four deformations as shown in  Fig.~\ref{fig:Fig8} for triangular and rhomboidal arrays. The plots are shown for $E=1.6 E_{b}$. The top-bottom asymmetry in panel a) is emphasized by the vertical displacement used in the figure, to allow for a direct comparison with panel b) where a fourth bubble is added at the bottom and the whole configuration is symmetric.

These structures, that exhibit signs of an emergent charge periodicity known as moir{\'e} patterns, give rise to several of the transport features studied in Refs.~\onlinecite{Zhang2018,Zhang2019}. 

In analogy with linear arrays, as the number of deformations increases, the regions with higher (lower) LDOS become better defined and the magnitude of the LDOS increases (decreases). The comparison with the linear array reveals that the magnitude for $\Delta \rho_j({\bf{r}},E)$ is larger in a triangular structure, a direct consequence of the simultaneous overlap of 3 regions with same pseudo-magnetic field signs. Notice that the magnitude decreases slightly when a fourth deformation is added as shown in panel b). This can be understood in terms of the picture of overlapping pseudo-magnetic regions, as the last deformation included at the bottom exhibits a field 'petal' with an opposite sign at the center. 

\begin{figure}
\includegraphics[width=0.48\textwidth]{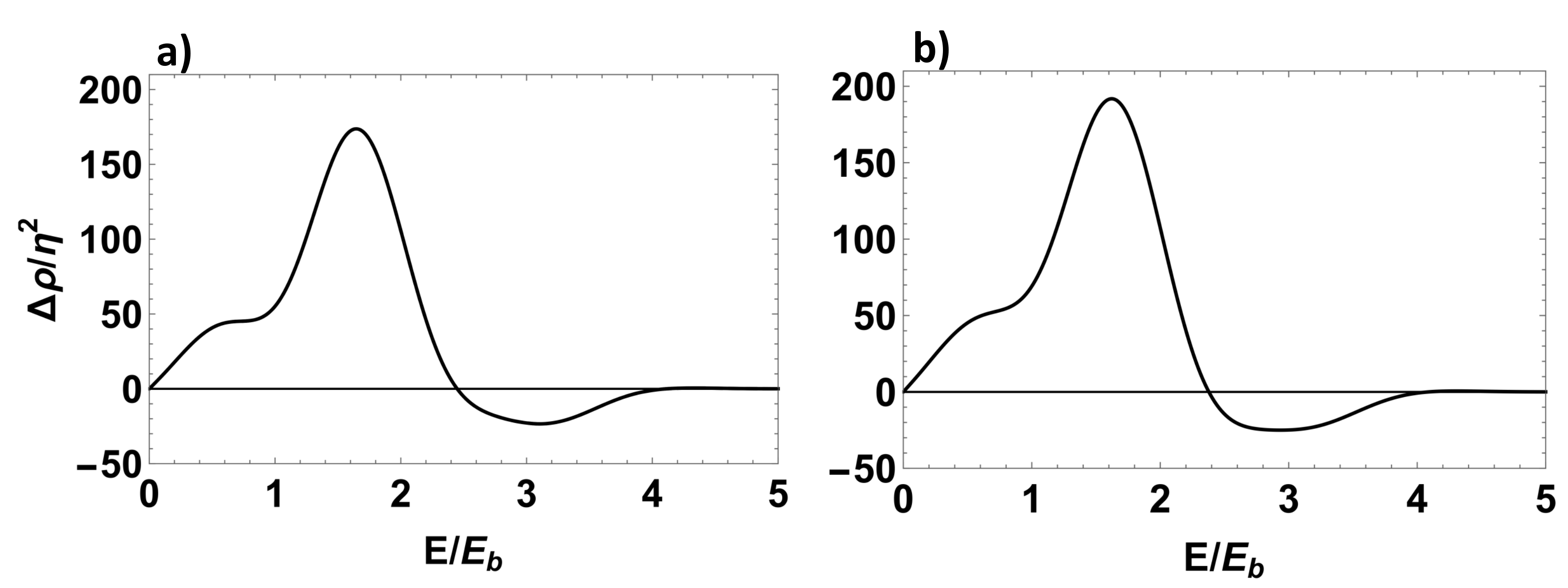}
\caption{$\Delta \rho_j({\bf{r}},E)$ induced by a triangular arrangement of Gaussian deformations centered at $(-b,0), (b, 0)$ and $(0,\sqrt{3}b)$ as function of energy for fixed positions marked in ~Fig.\ref{fig:Fig8}a). a) Variation shown for positions $(x = -b, y = 1.2b)$, and $(x = b, y = 1.2b)$ with perfect superposition between both curves. b) Variation shown for position $(x = 0, y = -0.6b)$. Units and other parameters as in Fig.~\ref{fig:Fig2}.}
\label{fig:Fig9}
\end{figure}

An analysis of the data for the three-bubble structure reveals distinct features that reflect the underlying lack of symmetry of this arrangement. Fig.~\ref{fig:Fig9} shows line cuts of Fig.~\ref{fig:Fig8}, at fixed positions symmetrically located at the centers of regions with maximum $\Delta \rho_j({\bf{r}},E)$ values. Panel a) reveal identical variations as a function of energy, at mirror symmetric points at each side of a vertical axis crossing through the origin (lines fall on top of each other for these two positions). Panel b) is a cut through a position along that axis $(x=0)$ and exhibits slight differences from the other two, consistent with non-identical surroundings.  

\begin{figure}
\includegraphics[width=0.45\textwidth]{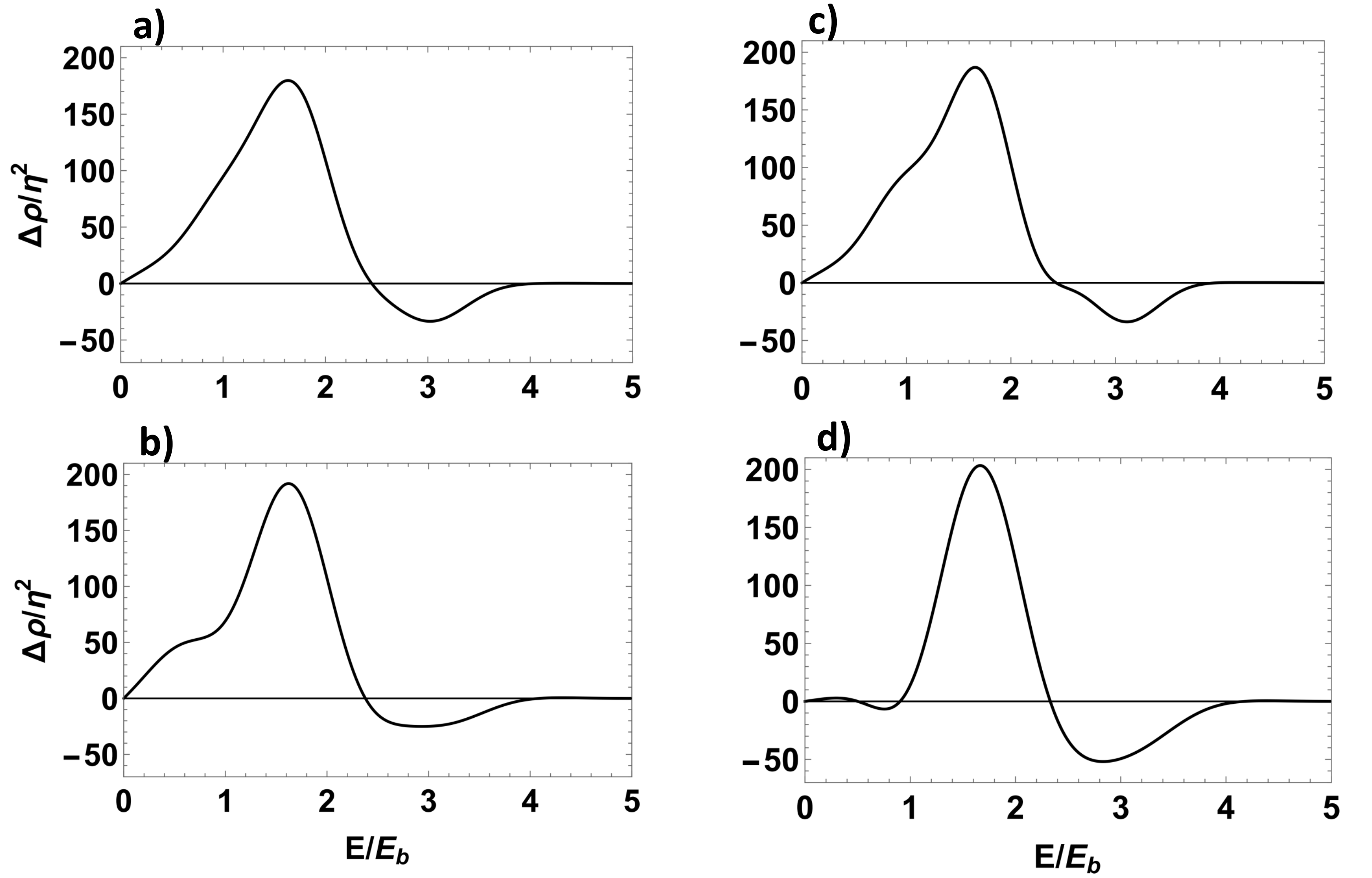}
\caption{$\Delta \rho_j({\bf{r}},E)$ for one- and two-dimensional arrays of 3 and 4 bubbles. Panel a) linear array of 3 bubbles. Panel b) triangular array of 3 bubbles. Panel c) linear array of 4 bubbles. Panel d) rhomboidal array of 4 bubbles. Units and other parameters as in Fig.~\ref{fig:Fig2}.}
\label{fig:Fig10}
\end{figure}

A contrast between one- and two-dimensional structures is shown in Fig.~\ref{fig:Fig10} for 3 and 4 bubbles. Top  panels show data for linear arrays while bottom panels show similar results for triangular and rhomboidal configurations. Because of the larger magnitudes and narrower peak structures two-dimensional arrays appear as better suited for designing enhanced charge accumulation regions. 

\begin{figure}
\includegraphics[width=0.49\textwidth]{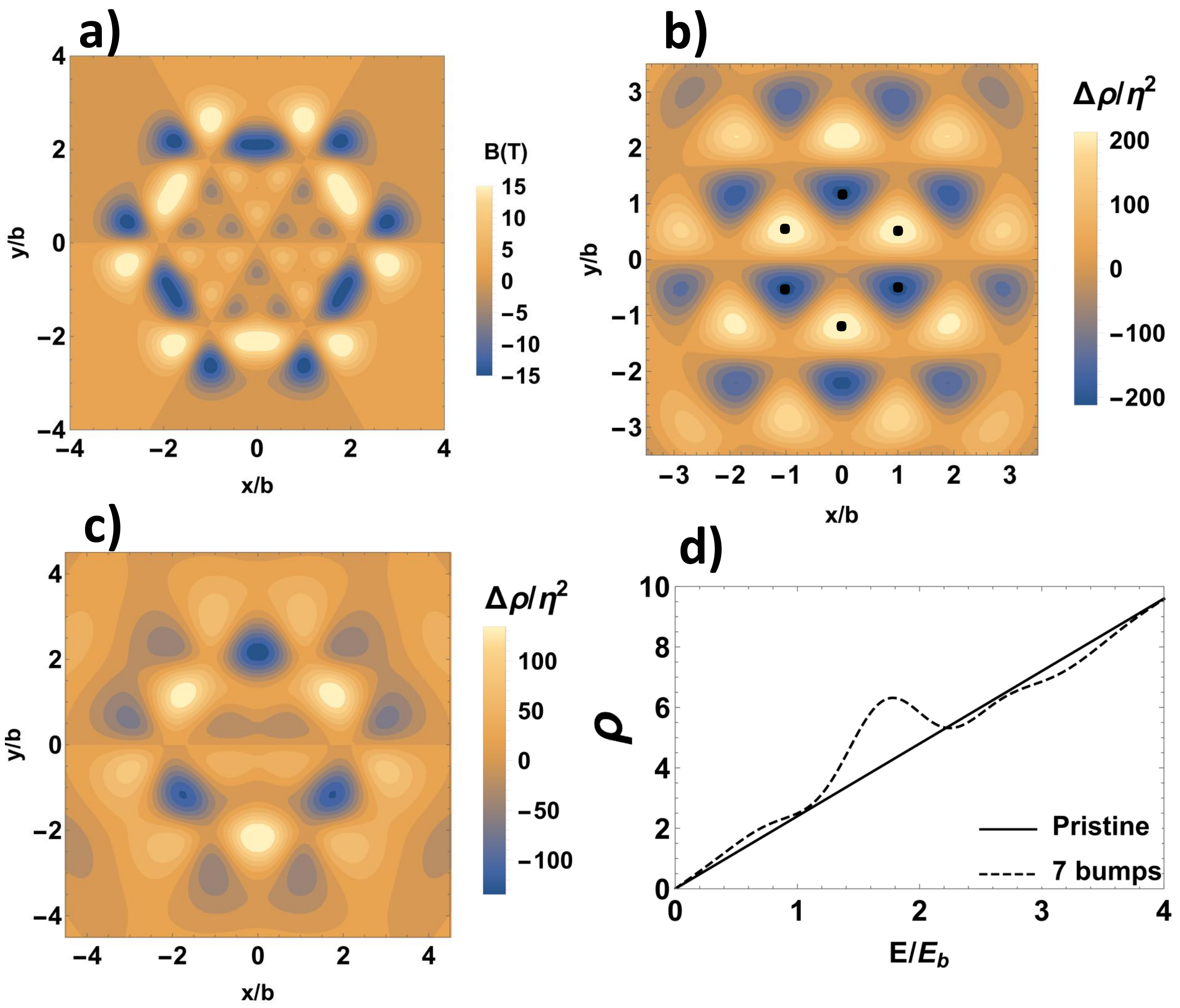}
\caption{Pseudomagnetic and $\Delta \rho({\bf{r}},E)$ profiles for a close-packed arrangement of 7 identical Gaussian deformations centered at $(0, 0)$, $(b, \sqrt{3}b)$, $(b, -\sqrt{3}b)$, $(-b, \sqrt{3}b)$, $(-b, -\sqrt{3}b)$, $(2b, 0)$, $(-2b, 0)$. Panel a) Pseudomagnetic field profile at valley $K$ (opposite signs/colors at valley $K'$, not shown). Panel b) $\Delta \rho({\bf{r}},E)$ profile at energy $E = 1.7 E_b$. Black dots indicate positions for Fig.~\ref{fig:Fig12}.Panel c) $\Delta \rho({\bf{r}},E)$ profile at energy $E = E_b$. Panel d) $ \rho({\bf{r}},E)$ at either position $(x = \pm 1, y = -0.6)$ or alternatively $(x =0, y= 1.2)$ within blue areas marked by black dots in panel b) (results for equivalent positions in yellow areas produce an inverted $ \rho({\bf{r}},E)$ not shown). Units and other parameters as in Fig.~\ref{fig:Fig2}.}
\label{fig:Fig11}
\end{figure}

Finally, we consider a close-packed structure of seven identical Gaussian deformations with center-to-center separation $s = 2b$. This particular arrangement is symmetric with respect to zigzag (x-axis) and armchair (y-axis) directions. 
Fig.~\ref{fig:Fig11} shows in panel a),  the profile produced by the corresponding pseudo-magnetic field, that serves as a guidance for the identification of the states contributing to the change in the LDOS. 
Two distinct regions can be clearly identified: Region 1) defined as the area inside the hexagonal profile with alternating triangular-shaped regions and pronounced pseudo-field gradients that act as non-constant magnetic barriers, and Region 2) defined by the boundaries of the hexagonal profile with extended areas that alternate between two roughly constant (average) values of pseudo-magnetic fields ($B_{av} \sim \pm 6.7$ T for the parameters chosen). 

\begin{figure}
\includegraphics[width=0.49\textwidth]{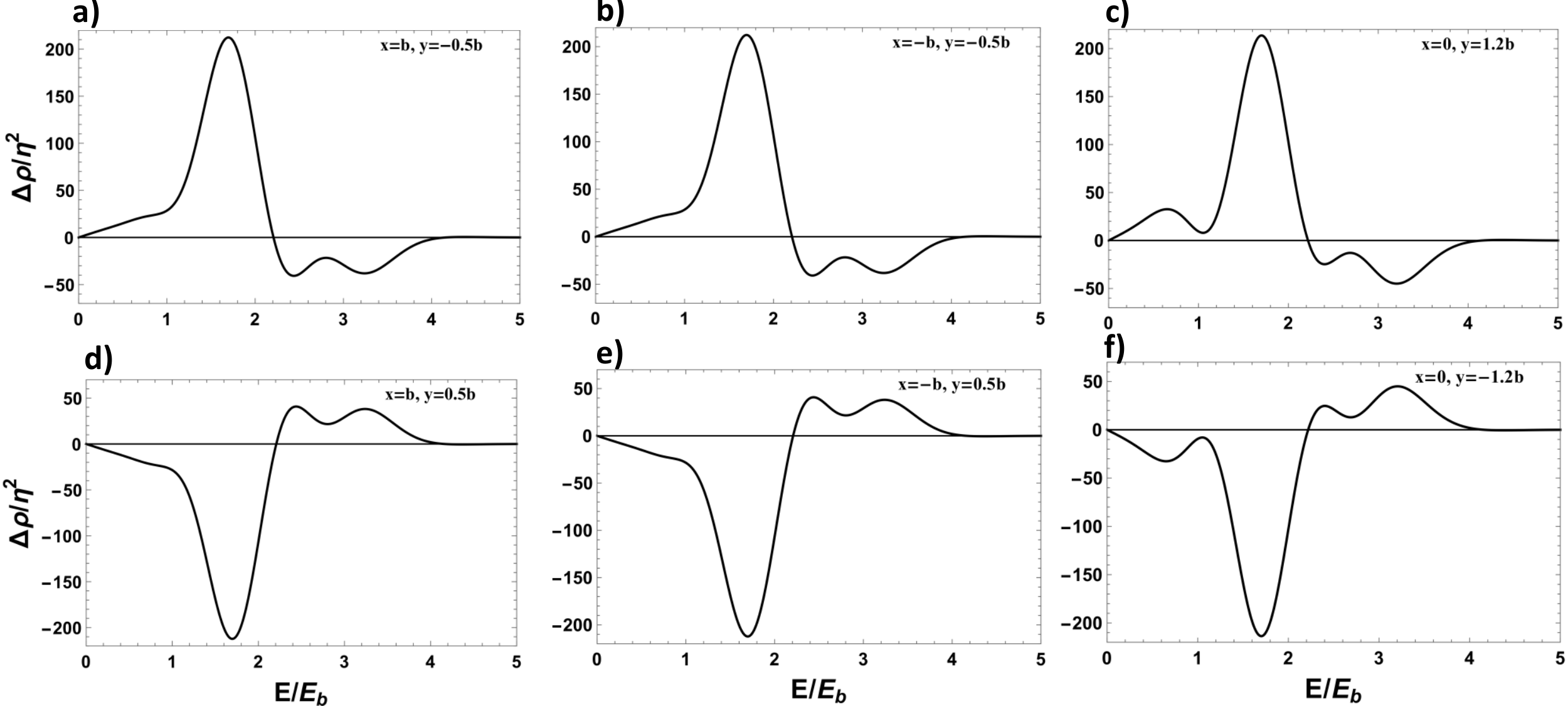}
\caption{$\Delta \rho_j({\bf{r}},E)$ for a close-packed arrangement of 7 identical Gaussian deformations centered at $(0, 0)$, $(b, \sqrt{3}b)$, $(b, -\sqrt{3}b)$, $(-b, \sqrt{3}b)$, $(-b, -\sqrt{3}b)$, $(2b, 0)$, $(-2b, 0)$ at positions shown in Fig.~\ref{fig:Fig11}. Units and other parameters as in Fig.~\ref{fig:Fig2}.}
\label{fig:Fig12}
\end{figure}

The structure in Region 1) is highly symmetric as shown by the spatial profile of $\Delta \rho_j({\bf{r}},E)$ in Panel b) of Fig.~\ref{fig:Fig11} and by the corresponding line cuts through the marked points shown in Fig.\ref{fig:Fig12}.
Peaks (dips) are better defined and exhibit larger amplitudes than previous structures (the small magnitude differences observed among the plots are numerical artifacts). 

Fig.~\ref{fig:Fig11} d) shows the corresponding LDOS as a function of energy at one of the three equivalent positions marked by black dots in the dark blue areas. The inverted curve is obtained for the corresponding bright yellow areas (note shown). In these internal regions, confined states arise due to the highly inhomogeneous pseudo-magnetic field profile \cite{Egger2007}. The corresponding energy level scaling is thus determined by the boundary conditions, i.e., by the shape and size of the confined region. For generic geometries, a simple argument can be made by assuming the size of these regions to be of order $b$, thus rendering $E \sim 1/b$\cite{Egger2007}. In a fixed strain configuration $B \propto \eta^2/b$ and one obtains $E \sim B$. (For the particular case of a single Gaussian deformation profile, the scaling in this regime has been discussed in Ref.~\onlinecite{DaweiMPLB}).The data shown in Fig.~\ref{fig:Fig11} b) suggests a constant separation between peaks (and/or dips) consistent with these ideas. For typical experimental settings as those in Ref.~\onlinecite{Zhang2019}, the energies of these 'bound' states estimated by this model are  $E = 1.7 E_b \simeq 60-70 meV$. Interestingly, this particular energy level scaling behavior has been reported in a recent publication on graphene samples undergoing a buckling transition that renders a periodic strain profile. Measurements in the region between contiguous corrugations appear to follow this linear scaling\cite{Mao2020}.

In contrast, Region 2) show areas with smoothly varying amplitudes for the pseudo magnetic field and provide more favorable conditions for the development of pseudo-Landau levels with magnetic length given by $l_B = \sqrt{\hbar/eB_{av}}$. These levels will exhibit an energy level scaling $E \sim \sqrt{B}$. This more common scaling regime has also been reported in Ref.~\onlinecite{Mao2020} for measurements on top of corrugated areas.

\section{Conclusion}

The study of a finite group of out-of-plane deformations in a graphene membrane reveals the emergence of periodic structures in the LDOS, reminiscent of moir{\'e} patterns observed in bilayer graphene and supported graphene on hBN substrates among other systems. The periodicity of these patterns is energy dependent and results from the underlying strain fields acting on the membrane. The inhomogeneities in the LDOS are stronger in areas where deformations overlap and their magnitudes can be directly related to the geometrical parameters characteristics of the underlying deformations.  The regions can take the shape of zero-dimensional areas -akin to quantum dots- or more extended, quasi-one dimensional structures as the energy is varied. In linear arrays, the quasi-one dimensional regions appear as waveguides for charge carriers and could be locally tuned with appropriate changes in chemical potentials.
Two-dimensional close-packed structures also can exhibit enhanced LDOS at discrete energies giving rise to pseudo-Landau levels and pseudo-magnetically confined states. Depending on the geometrical parameters defining these regions either or both types of states could be observed, with the energy scaling showing the corresponding crossover between $E \propto \sqrt{B}$ and $E \propto B$ regimes. 
The emergence of elements of moir{\'e} patterns described in this paper, suggest that by choosing appropriate substrates it is possible to create extended regions with multiple separated areas of enhanced LDOS, akin to quantum dot arrays by design. The quantitative connection established between geometrical parameters of deformations and charge distributions reported in this work, provides the necessary tools to carry out such program. On  a final,  more speculative note, it is interesting to consider the design of structures able to confine only two-levels in each internal area.  As these levels are separated by pseudo-magnetic barriers, they would present a realization of two-level systems or qubits, that could be addressed externally by local probes. The investigation of such a regime is left for a future report.

\begin{acknowledgments}
We acknowledge discussions with D. Zhai, Avalos-Ovando, D. Faria, H. E. Castillo and S.E. Ulloa. This work was partially supported by NQPI-Ohio University. Portions of this work were completed at the Aspen Center for Physics under support from NSF grant No.PHY-1607611.
\end{acknowledgments}

\bibliography{MoireRefs}
\bibliographystyle{apsrev4-2}
\end{document}